# Measurement of the Range Component Directional Signature in a DRIFT-II Detector using $^{252}$Cf Neutrons

S. Burgos [a], E. Daw [b], J. Forbes [a], C. Ghag [c], M. Gold [d], C. Hagemann [d], V.A. Kudryavtsev [b], T.B. Lawson [b], D. Loomba [d], P. Majewski [b], D. Muna [b], A. St.J. Murphy [c], G.G. Nicklin [b], S.M. Paling [b], A. Petkov [a], S.J.S. Plank [c], M. Robinson [b], N. Sanghi [d], D.P. Snowden-Ifft [a][*], N.J.C. Spooner [b], J. Turk [d] and E. Tziaferi [b]

*Corresponding author.

*Telephone*: (323) 259-2793

*FAX*: (323) 341-4858

*E-mail address*: ifft@oxy.edu (D.P. Snowden-Ifft)

[a] *Department of Physics, Occidental College, Los Angeles, CA 90041, USA.*

[b] *Department of Physics and Astronomy, University of Sheffield, Sheffield, S3 7RH, UK.*

[c] *School of Physics, University of Edinburgh, Edinburgh, EH9 3JZ, UK.*

[d] *Department of Physics and Astronomy, University of New Mexico, NM 87131, USA.*

## Abstract

The DRIFT collaboration utilizes low pressure gaseous detectors to search for WIMP dark matter with directional signatures. A $^{252}$Cf neutron source was placed on each of the principal axes of a DRIFT detector in order to test its ability to measure directional signatures from the three components of very low energy (~keV/amu) recoil ranges. A high trigger threshold and the event selection procedure ensured that only sulfur recoils were analyzed. Sulfur recoils produced in the $CS_2$ target gas by the $^{252}$Cf source closely match those expected from massive WIMP induced sulfur recoils. For each orientation of the source a directional signal from the range components was observed, indicating that the detector is directional along all 3 axes. An analysis of these results yields an optimal orientation for DRIFT detectors when searching for a directional signature from WIMPs. Additional energy dependent information is provided to aid in understanding this effect.

## Introduction

The identification of dark matter remains one the outstanding goals of contemporary physics. Weakly Interacting Massive Particles (WIMPs) are a strongly motivated candidate [1]. Over the last few decades these facts have motivated a large number of researchers worldwide to attempt to detect these elusive particles. One group, [2, 3], claims to have detected the annual modulation signature from WIMPs. Though controversial, this result does highlight the need for an unambiguous, robust signature from WIMPs. Directional signatures of WIMPs provide [4, 5, 6] the most robust signatures of WIMP interactions. The Directional Recoil Identification From Tracks (DRIFT) collaboration utilizes low pressure negative ion TPCs specifically designed to measure directional signatures associated with WIMP interactions within it. Several ~1 m$^3$ DRIFT detectors have run for extended periods of time in the Boulby Mine in England [7]. This and two associated papers explore the ability of DRIFT detectors to measure directional signals from WIMPs.

As discussed in [4, 5, 6] the Earth is thought to move through an essentially stationary halo of WIMPs surrounding our Galaxy. As such a WIMP wind is created coming from the direction of the constellation Cygnus as viewed from an observer on the earth. The declination of Cygnus is ~45 degrees. A directional detector located at a latitude of 45 degrees will therefore see this WIMP wind vector oscillate, over a sidereal day, from pointing towards the center of the Earth to pointing south, see Figure 1 where these directions are used to define the *x* and *z* axes respectively. This oscillation occurs every sidereal day and will rapidly go out of phase with a terrestrial day over the course of a year long run. Backgrounds that may have a natural modulation period of a terrestrial day can therefore be separated from WIMP recoils. Recoil ranges produced by WIMPs within the detector and projected onto the *x-z* plane will preferentially lie in the same direction as the WIMP "wind" vector [8]. The components of the range parallel to these directions, $\Delta x$ and $\Delta z$, will therefore oscillate, 90 degrees out of phase, over a sidereal day. A number of effects conspire to dampen the magnitude of these oscillations; WIMPs are neither mono-energetic nor uni-directional; recoils from WIMPs are also neither mono-energetic nor uni-directional; recoil energies (~1 keV/amu) are small and therefore the recoils do not travel in straight lines; ionization from recoils diffuses on the way to the detector; and finally resolution issues associated with any measurement of the magnitude of a component of the range can also cause the magnitude of these range component oscillations to be smaller than expected. The DRIFT-II detectors measure all 3 components of the range of the recoils. To determine how well the DRIFT detectors are able to reconstruct the three orthogonal range components, a $^{252}$Cf neutron source was placed on each of the *x*, *y* and *z* axes to produce recoils inside of one of these detectors. Data taken from these runs was then analyzed to estimate the magnitude of the expected oscillations.

## Experimental Setup

The design of DRIFT-II detector modules has been described in detail elsewhere [9]. The module employed in this study, known as DRIFT-IIc, is located in a surface laboratory at Occidental College, Los Angeles, CA, U.S.A. A brief description of the detector relevant to this experiment is presented here for convenience. A $1.5^3$ $m^3$ low background stainless steel vacuum vessel provided containment for 40 Torr of $CS_2$. The $CS_2$ gas, supplied by evaporation from the liquid state, flowed through the vessel steadily by means of a mass flow controller at the input and a dry pump at the output. Within the vacuum vessel were two back-to-back TPCs with a shared, vertical, central cathode constructed of 20 μm stainless steel wires with 2 mm pitch, see Figure 1. Two field cages, located on either side of the central cathode, defined two drift regions of 50 cm depth in which recoil events could form. Charge readout of tracks was provided by two MWPCs each comprised of an anode plane of 20 μm stainless steel wires with 2 mm pitch sandwiched between (1 cm gap) two perpendicular grid planes of 100 μm stainless steel wires also with 2 mm pitch. The potential difference between the grids and the grounded anode planes was -2885 V. The central cathode voltage at -32 kV produced a drift field of 587 V/cm. For each MWPC, 448 grid wires ($y$-direction) were grouped down to 8 "lines" which were then pre-amplified, shaped and digitized. The anode wire signals ($x$-direction) were treated identically except that the electronic gain was half that of the grids. Eight adjacent readout lines (either anode or grid) therefore sampled a distance of 16 mm in $x$ and $y$. Voltages on the grid and anode lines were sampled at 1 MHz providing information about the event in the third ($z$-direction) dimension. The 52 wires at the edges of the grid and anode planes were grouped together to provide veto signals for each MWPC. Triggering of the data acquisition system (DAQ) occurred on individual anode lines for this experiment at a level of 200 ADC counts. All lines were digitized from -3000 μS to +7000 μS relative to the trigger with 12 bit digitizers. The region bounded by the vetoes and the inner grid planes formed a fiducial volume of 0.80 m$^3$ equating to a 134 g mass of $CS_2$ within which recoil events could occur. Each side of the detector was instrumented with an automated, retractable, ~100 μCi $^{55}$Fe calibration sources which allowed monitoring of detector gain and functionality.

In order to test for directional signatures a 202 μCi $^{252}$Cf neutron source was placed on each of the DRIFT-IIc's principal axes (*x*, *y* and *z* as described above). The source was moved 331 cm away from the center of the detector so that the angular spread of the neutrons interacting in the detector (excluding scattering) was ~10 degrees. Data were taken in which the neutrons were directed, on average, in the +*x*, -*y*, +*z* and –*z* directions. More than 50,000 events were recorded to disk in each of these orientations. Background (no source present) and $^{55}$Fe source calibration data were also recorded. Figure 2 shows an event from a +*x* run.

## Comparison of $^{252}$Cf Neutron Recoils with Expected WIMP Recoil Events

To test the suitability of $^{252}$Cf neutrons for simulating the interactions of WIMPs, the Monte Carlo codes *GEANT 4* [10] and *Cygnus* [8] were used to generate simulated recoils from neutrons and simulated recoils from massive WIMPs. *GEANT4*, which was successfully compared to data from the DRIFT-IIa detector [7], was modified to simulate the laboratory at Occidental College. Approximately 30% of the neutrons interacting in the detector were found to have scattered (in the floor, vessel, detector etc.) and entered the detector at angles larger than the beam spread, ~10 degrees.  A program, called *Cygnus*, and discussed in [8], generated simulated WIMP recoils.  Figure 3 shows the sulfur *NIPs* (for number of ion pairs) spectrum from $^{252}$Cf neutrons and from 1000 GeV/c^2 WIMPs generated by these two simulations.  Figure 4 shows the simulated spectrum of recoil angles from $^{252}$Cf neutrons from a –z run.  Figure 4 also shows the simulated spectrum of recoils angles from 1000 GeV/c$^2$ WIMPs when the direction of the WIMP "wind" is pointed in the –z direction.  The $^{252}$Cf neutrons produce a harder recoil spectrum but the WIMP recoils are more forward peaked.  For a directional experiment these are offsetting effects.  Therefore sulfur recoils produced by the $^{252}$Cf source in this experiment provide a reasonable facsimile of sulfur recoils produced from massive WIMPs.  The same cannot be said about carbon recoils.  Carbon recoils from $^{252}$Cf neutrons have much higher energies than from massive WIMPs.

## Event Selection and Analysis

The background trigger rate was ~2 Hz, rising to ~15 Hz with the source in place.  Background triggers due to sparks, uncontained events, alpha background and ringers [11] were removed using cuts on these types of events.

The selection of a software threshold for analysis was the first cut on the data.  This was chosen to be 50 ADC counts or ¼ of the hardware trigger level of the DAQ in order to include ionization on neighboring wires for the anode.  For the veto and the grid the thresholds were set to 30 and 100 ADC counts respectively.  Any line that was not triggered was not analyzed further.  For large events, the induced pulses on neighboring wires were sometimes big enough to trigger the analysis so a cut was introduced to remove these lines from further analysis.  An example is shown on the right yellow anode line (6[th] from the bottom) in Figure 2. Each line that passed these cuts was considered a "hit" and analyzed further.

A region of interest (ROI) was defined as extending from 200 μS before the trigger time to 500 μS after the trigger time in which 22 statistics were calculated for each hit. These statistics are discussed in detail in [11]. With these statistics a set of simple cuts were placed on the data to ensure that the ionization could be properly calculated. These include cuts to ensure that the digitizers did not saturate, that the ionization fell within the ROI on both the anode and the grid and that it was contained inside the detector's fiducial volume (i.e. no signals from the veto regions of the detector).

One of the most important parameters associated with each event was the total amount of ionization, *NIPs*. As discussed in [7] $^{55}$Fe calibrations on both sides of the detector allowed for the conversion of the integral of the voltage with time, on both the grid and the anode, into *NIPs*. An average of all of the $^{55}$Fe results over the runs provided the best estimate of this conversion parameter. An anomalous population (~10% of the total) of events passing all of the above cuts was observed to have significantly less *NIPs* on the grid than on the anode. These were removed.

In preparation for the spatial analysis several more cuts were placed on events. Any event with 8 hits was cut. Lines that triggered the analysis were required to be adjacent to each other. Events were required to appear on only one side of the detector. Events occurring inside the MWPCs were removed with a cut on the full-width-at-half-maximum (*FWHM*) of individual lines. Though ringers [7, 10] made up only a small fraction of the events recorded to disk a cut was placed on them using the derivative of a smoothed waveform (20 μS smoothing and differentiating time). This methodology proved simpler and more robust than that discussed in [11].

With all cuts in place the accepted event rates for the background runs dropped from ~2 Hz to ~0.01 Hz while the event rates for the neutron runs dropped from ~15 Hz to ~2 Hz. Column 2 of Table 1 shows the number of events passing all of the above cuts for each of 3 neutron source orientations *(+x, -y and +z)*. Figure 5 shows a histogram of the *NIPs* obtained with the aforementioned cuts. As can be seen in this plot the hardware threshold of the DAQ and the analysis cuts produced an effective threshold of ~1000 *NIPs*. A 1000 *NIPs* sulfur(carbon) recoil has an energy of 47.2(30.9) keV [11].

The length of the track in *x*, $\Delta x$, was calculated in the following way. The number of anode wires that were hit was multiplied by 2 mm, the inter-wire spacing, and 1 mm subtracted from the total. For $\Delta z$ the following procedure was adopted. The $\Delta z$ parameter is related to the time duration of the deposition of charge onto the wires. Ideally the length of an event in time would be calculated by utilizing the sum of all 8 anode lines. Unfortunately induced pulses, see Figure 2, severely affect this summed pulse. Due to the grouping of the wires the induced pulses on anode lines far from the main ionization have about the same size. Therefore the line furthest from the line with the most ionization was subtracted from all of the other lines. Lines which were originally above the software threshold, that is hits, were then summed together to make a combined waveform. The *FWHM* of this summed pulse was then calculated. $\Delta z$ was then calculated by multiplying the drift velocity, $v_{drift}$ = 5032 cm /s, by this *FWHM*. Finally $\Delta y$ was obtained in the following way. For each 1 μS time interval within the ROI the grid line with the lowest voltage was subtracted from the rest. A Gaussian was then fit to all of the 8 data points and the width and centroid of that fit recorded. From the geometry of the detector a width of 2 mm [12] was expected. As a goodness of fit criterion, any time with a fit width of 4 mm or less was kept for further analysis. These selected times were multiplied by $v_{drift}$, producing $z$ values, and plotted against the selected centroids, average $y$s. A line was fit to these data. The slope, $\Delta y/\Delta z$, of this fit was then multiplied by $\Delta z$, from above, to obtain $\Delta y$. Finally because $\Delta y$ cannot be accurately measured unless the track is inclined with respect to the detector it was required that $\Delta z > 0.22$ cm. This removed 70% of the events from $\Delta y$ analysis for the $+x$ and $-y$ runs and 60% of the events from the $+z$ run.

## Results

Figure 6 shows the distributions for range components $x$, $y$ and $z$ for each of the 3 source orientations. It is interesting to note that while the distributions do show differences, due to the different ways in which $\Delta x$, $\Delta y$ and $\Delta z$ were calculated, none shows a significant population of track lengths above 1 cm in length. While this is consistent with the lengths of tracks expected from sulfur recoils, it suggests that the energy deposition rate of carbon recoils is too low to result in enough ionization density on a single wire high enough to trigger the data acquisition. This behavior is identical to that which causes the very high gamma ray rejection achieved by DRIFT [7]. Comparison of the $\Delta z$ data (considered to be the most reliable) to a GEANT4 simulation suggests an insensitivity to carbon recoils of at least a factor of 100. Consequently, the results presented in this paper can be considered to apply only to sulfur recoils.

Table 1 shows the means of $\Delta x$, $\Delta y$ and $\Delta z$ for each of the 3 neutron source orientations for all of the events passing the event selection criteria. Because of the different ways in which $\Delta x$, $\Delta y$ and $\Delta z$ were calculated a comparison of, for instance, $\Delta x$ to $\Delta y$ is not possible. It is possible, however, to compare the mean $\Delta x$ from one orientation of the source to the mean $\Delta x$ from another orientation of the source. Table 1 reveals that for each parameter, $\Delta x$, $\Delta y$ and $\Delta z$, there is a direction for the neutrons which produces the largest mean. The mean $\Delta x$ is largest when the source is on the *x* axis, the mean $\Delta y$ is largest when the source is on the *y* axis and the mean $\Delta z$ is largest when the source is on the *z* axis. This was previously observed in [11] and is predicted by simulations. Figure 7 shows the data in Table 1 graphically. The range components ($\Delta x$, $\Delta y$ and $\Delta z$) are shown on the abscissa. The ordinate shows the difference from the mean of the two perpendicular directions. The data from perpendicular directions are consistent with each other though there is some evidence for systematic effects. The data from parallel directions are significantly different than the perpendicular directions indicating a sensitivity to the presence of directed neutrons. Given the similarity between the neutron and WIMP recoil spectra, shown in Figures 3 and 4, it is reasonable to assume that the same will hold true for massive WIMPs. Further experiments and simulation work will make this case more quantitative.

## Discussion

Clearly $\Delta y$ is not as effective as $\Delta x$ or $\Delta z$. Reasons for this include the loss of events caused by the cut on $\Delta z$, small signal to noise caused the generation of the induced pulses which are spread over all wires and a compression the ionization in the *y* direction caused by the electric field lines being compressed between the grid wires. This compression is the reason why the mean $\Delta y$ is small compared to $\Delta x$ and $\Delta z$. As expected $\Delta z$ is DRIFT-IIc's best directional parameter. This is due to the high resolution associated with this measurement.

As discussed in the introduction a detector located at a latitude of ~45 degrees and properly oriented would "see" the constellation Cygnus oscillate between being overhead and being on the northern horizon. $\Delta x$ and $\Delta z$ provide the most robust directional signatures. Therefore in order to search most effectively for WIMP recoils DRIFT detectors should be oriented with one of these axes along a North-South line and one vertical. Figure 1 shows a possible orientation. In this orientation the parameter $\Delta z/\Delta x$ would modulate over a sidereal day from a maximum, when the constellation Cygnus lies near the $z$ axis (overhead), to a minimum when the constellation Cygnus lies along the $x$ axis (on the northern horizon). The magnitude of this oscillation can be estimated by calculating $\Delta z/\Delta x$ from the $x$ and $z$ orientations of the neutron source. This is shown in the last column of Table 1. $\Delta z/\Delta x$ from $z$ directed neutrons is larger than $\Delta z/\Delta x$ from -$x$ directed neutrons by 12 standard deviations. Because of the different ways that $\Delta x$ and $\Delta z$ are defined this ratio does not oscillate around 1. As a check this ratio is also calculated for the -$y$ oriented neutrons and shown in Table 1. The value for $\Delta z/\Delta x$ from –$y$ directed neutrons falls midway between that for the -$x$ and +$z$ directed neutrons. A 14% oscillation can be expected with this parameter from WIMP recoils.

These results apply for sulfur recoils with a *NIPs* distribution shown in Figure 5. In order to aid in understanding this effect Figure 8 shows the amplitude of the oscillation, expressed as a percentage of the mean value, as a function of *NIPs*. $\Delta z/\Delta x$ is found to have a strong dependence on the recoil energy associated with the event. This is to be expected, since events with higher energy will be longer and therefore better able to distinguish themselves in the presence of diffusion in the drift volume. Two notable features of this plot are the linearity with the recoil energy, and the disappearance of the effect at 1000 *NIPs*, or 47.2keV sulfur recoil energy. With the detector readout, cuts and analysis used in this paper, 47.2 keV may be taken as a threshold for the observation of this directional signature.

## Future Work

The results in the current paper are gratifying in that they demonstrate, for the first time, significant directional sensitivity in a large dark matter detector. They should not be interpreted as the ultimate directional sensitivity of the DRIFT design. The $\Delta z/\Delta x$ can be made more powerful by improved reconstruction of the $\Delta z$ component, realized through higher signal to noise ratio and less aggressive pulse shaping in the post-charge-amplifier electronics. Lower noise in the grid readout could also be achieved, so that more sensitive measures of directionality could be constructed using information from the $\Delta y$ component of the event range. Revised readout electronics may be able to address this issue also. Testing and development of prototype electronics aimed at addressing these issues is underway. Furthermore reduction of diffusion through either higher drift fields or smaller drift lengths will allow better directional signals and lower energy thresholds. Finally the ability to measure the $z$ coordinate of the events would allow for a statistical subtraction of the diffusion.

## Conclusion

A $^{252}$Cf placed on the principal (*x*, *y* and *z*) axes of the DRIFT-IIc detector was used to produce recoils with directional signatures. A high trigger threshold and analysis cuts ensured that only sulfur recoils were recorded and made it through the event selection process. Measurements of the 3 range components *Δx*, *Δy* and *Δz* were made for each event. The average of these components was found to be significantly greater when the source was placed on the corresponding axis proving that a DRIFT detector can sense the direction of recoiling sulfur atoms. This is relevant to dark matter searches with DRIFT detectors because the energy and angular spectrum of simulated $^{252}$Cf and massive WIMP exposures are similar. The $^{252}$Cf exposures along the axes of the detector were additionally useful because, for detectors located at ~45$^0$ latitude, the average WIMP velocity vector points, predominantly, along 2 axes. A DRIFT detector located at ~45$^0$ latitude could expect an oscillation in the variable *Δz*/*Δx* of magnitude 14% with a threshold of 47.2 keV. The magnitude of this oscillation grows with energy. Additionally improvements to the electronic readout and to the design of the detector could improve the magnitude of this oscillation.

## Acknowledgments

We acknowledge the support of the US National Science Foundation (NSF). This material is based upon work supported by the NSF under Grants Nos. 0600840, 0600789 and 0548208. Any opinions, findings, and conclusions or recommendations expressed in this material are those of the author(s) and do not necessarily reflect the views of the NSF. We also acknowledge support from the Science and Technology Facilities Council (UK), Cleveland Potash Ltd., the European Commissions's FP6 ILIAS contract number RII3-CT-2004-506222 and the European Union Marie Curie International Reintegration Grant number FP6-006651.

**Tables**

Table 1

| Direction of neutrons | N | $\Delta x$ | $\Delta y$ | $\Delta z$ | $\Delta z/\Delta x$ |
|---|---|---|---|---|---|
| +x | 8673 | 0.285 +/- 0.002 | 0.116 +/- 0.002 | 0.2128 +/- 0.0009 | 0.964 +/- 0.006 |
| -y | 5859 | 0.261 +/- 0.002 | 0.129 +/- 0.003 | 0.209 +/- 0.001 | 1.019 +/- 0.008 |
| +z | 5829 | 0.270 +/- 0.002 | 0.121 +/- 0.003 | 0.233 +/- 0.001 | 1.092 +/- 0.009 |

# Figures

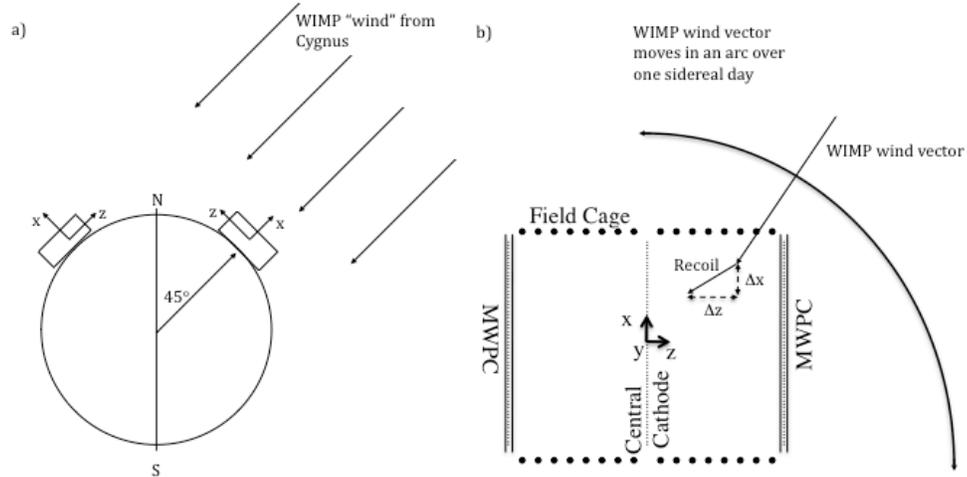

Figure 1 – This figure shows the motion of the WIMP "wind", coming approximately from the constellation Cygnus, in the Earth's reference frame. Under the simplifying assumption that Cygnus is at a declination of 45 degrees, a detector located at a latitude of 45 degrees would see the WIMP "wind" vector oscillate from horizontal (pointing from the north to the south) to vertical (pointing towards the center of the earth) over a sidereal day.

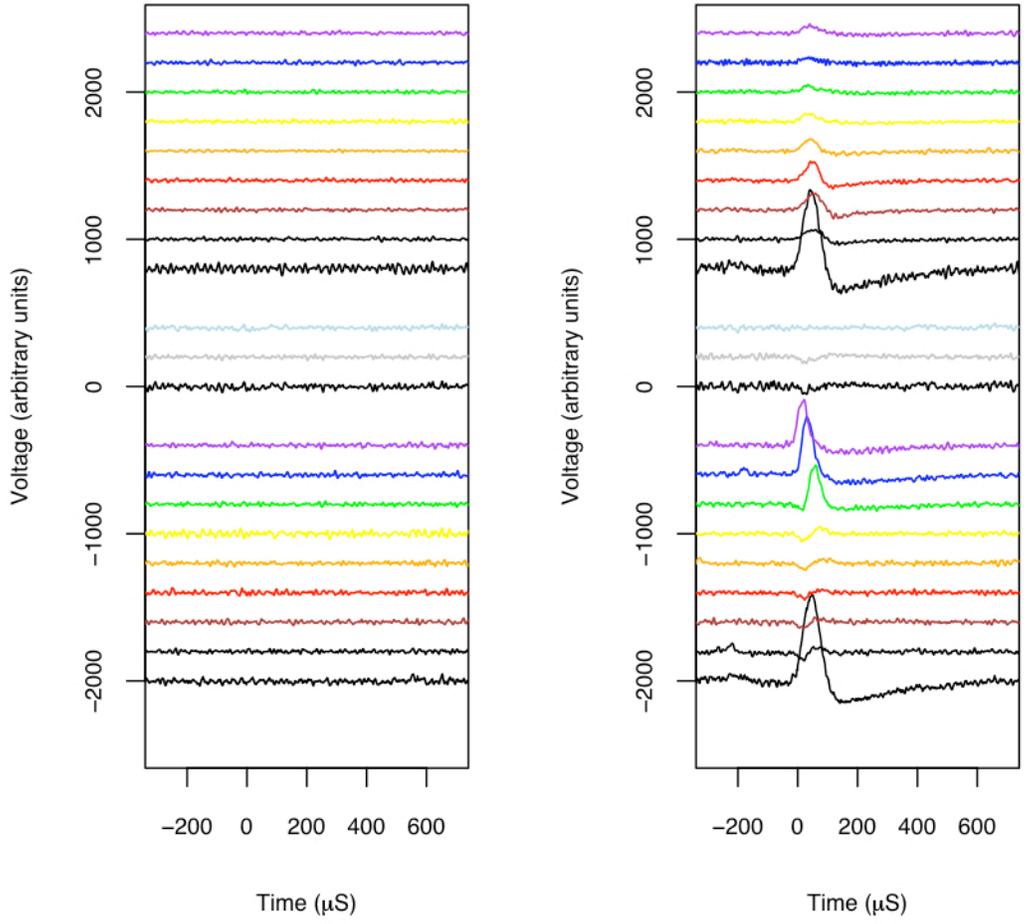

Figure 2 – An event from the *+x* neutron run. The left and right plots show data taken from the left and right MWPCs respectively. As discussed in the text, the 448 anode and grid wires were grouped down to 8 lines for digitization. The 8 digitized lines from the grid are shown at the top while the 8 digitized lines from the anode are shown to the bottom. The waveform presented below each set of 8 lines is the sum line. Unlike previous publications of DRIFT data the polarities of the lines have been reversed so that ionization falling on a wire, whether on the grid or anode, causes the signal to go positive. Shown between the anode and grid lines are the veto lines. The anode veto line is shown at about 200 on the vertical scale in light blue in the color version while the grid veto line is shown at bout 400 on the vertical scale in grey in the color version.

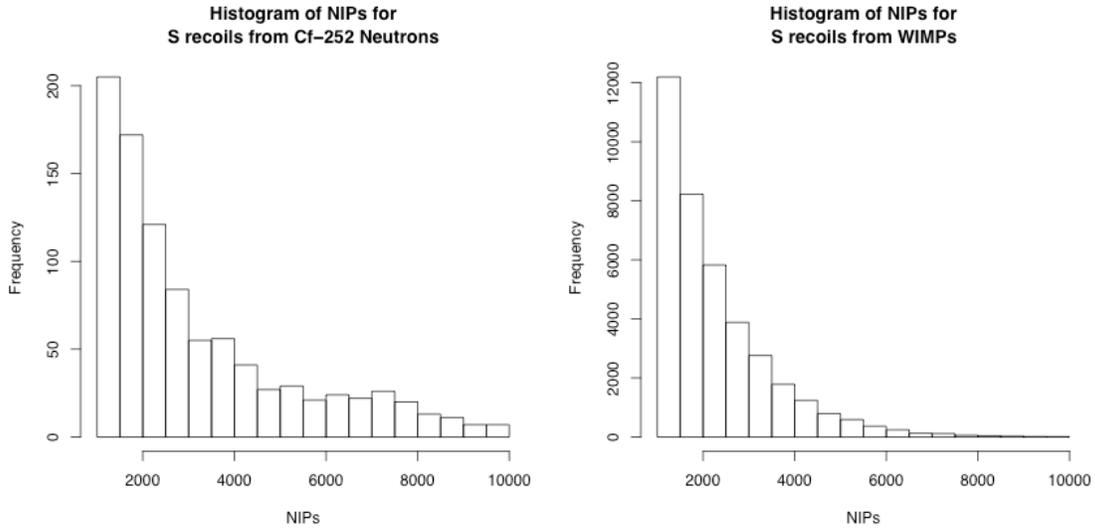

Figure 3 – The histogram on the left shows the *GEANT4* predicted spectrum of *NIPs* produced by sulfur recoils from an exposure to $^{252}$Cf neutrons. The histogram on the right shows the predicted spectrum of *NIPs* produced by sulfur recoils from an exposure of $CS_2$ to 1000 GeV WIMPs calculated using the Monte Carlo *Cygnus* discussed in the text.

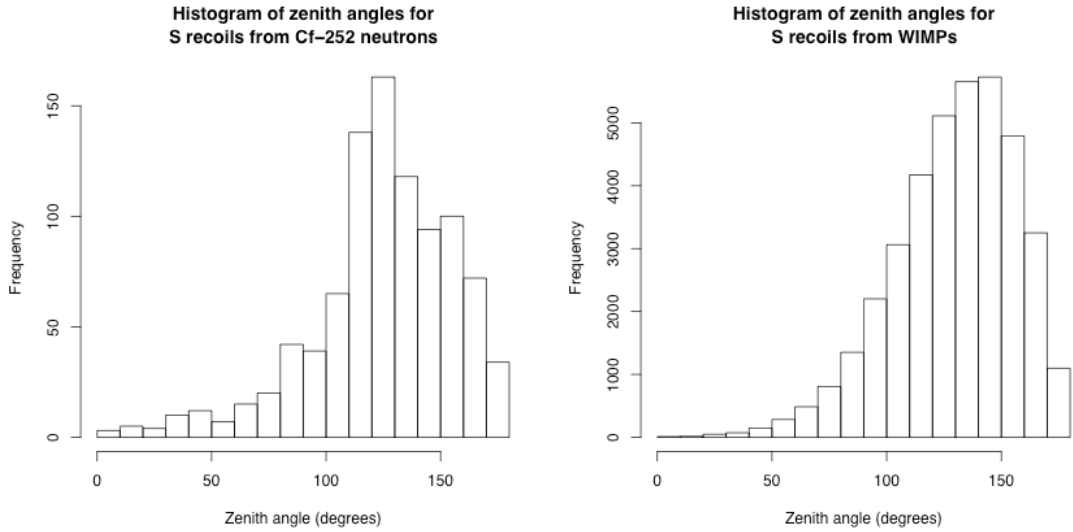

Figure 4 – The histogram on the left shows the *GEANT4* predicted spectrum of sulfur recoil zenith angles (measured from the *z* axis) from an exposure to $^{252}$Cf neutrons primarily directed in the –z direction. The histogram on the right shows the spectrum of sulfur recoil angles from an exposure to 1000 GeV WIMPs where the direction of the WIMP "wind" points in the –z direction. The latter is predicted by the program *Cygnus* discussed in the text. Only sulfur recoils producing 1000 - 10000 *NIPs* were included in these histograms. As can be seen on these plots the WIMP recoils are peaked at higher values of $\theta$ than the $^{252}$Cf recoils. This is due to the fact that neutrons preferentially forward scatter while WIMPs would produce an isotropic recoil spectrum.

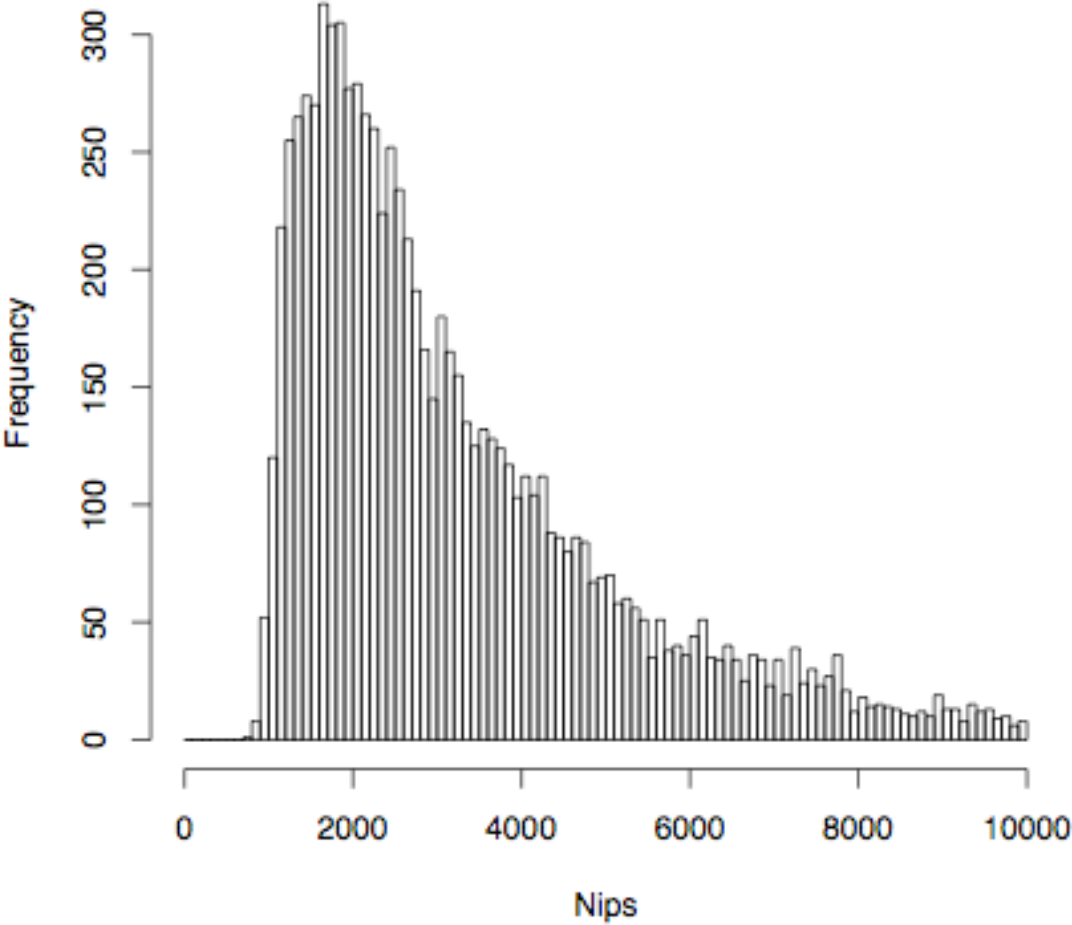

Figure 5 - The distribution of *NIPs* for selected events.

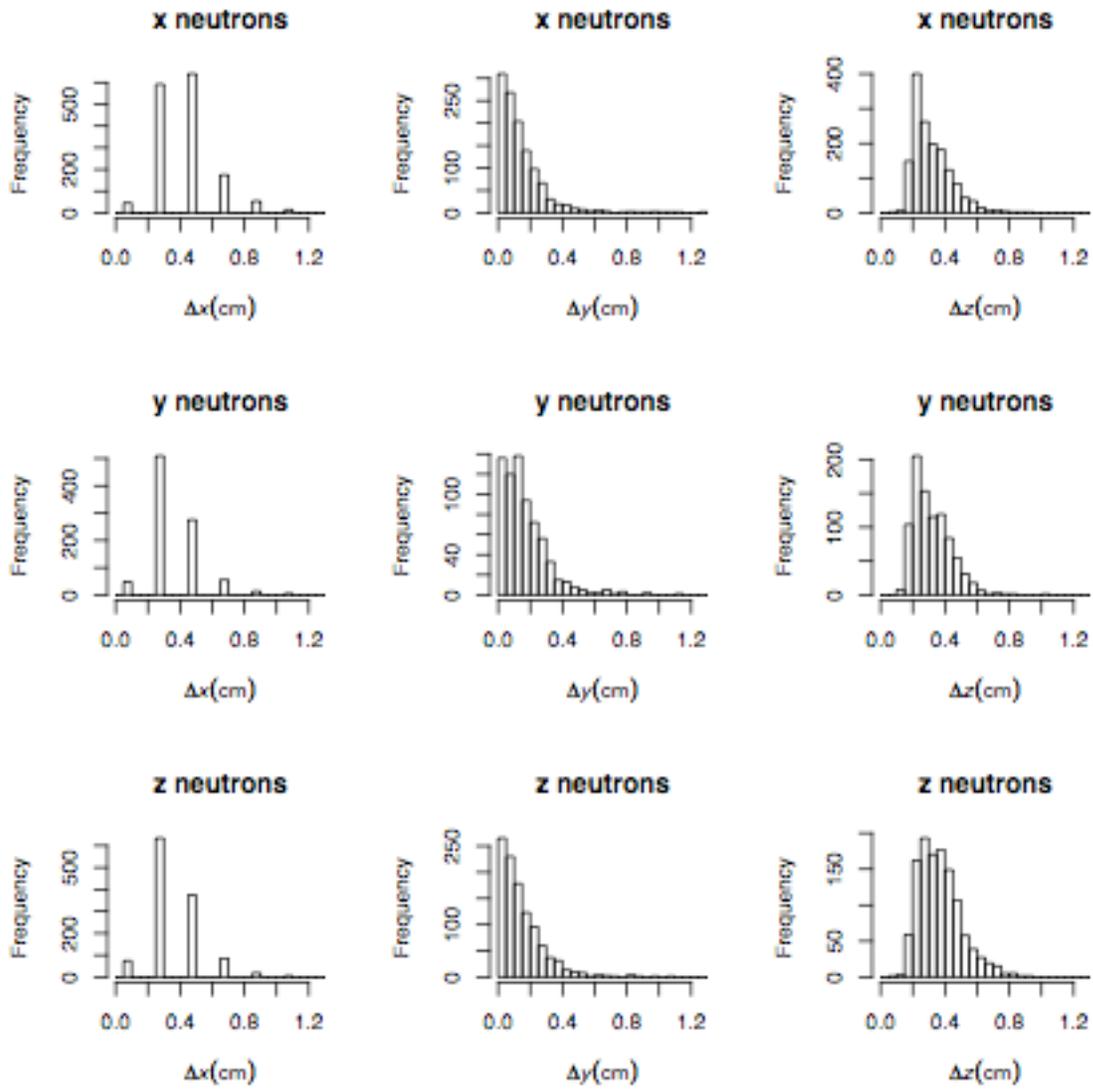

Figure 6 – These figures show the distributions of the various measured parameters ($\Delta x$, $\Delta y$ and $\Delta z$) for the *x*, -*y*, and +*z* neutron runs for events with more than 5000 *NIPs*.

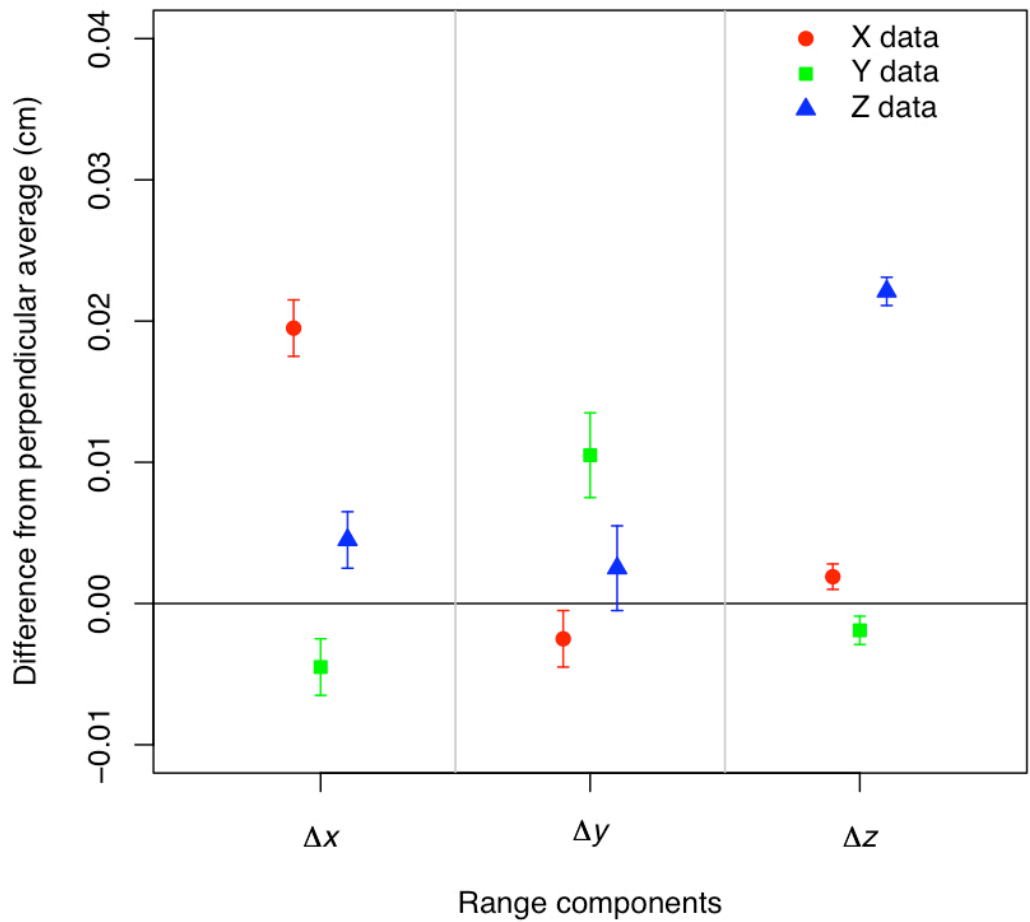

Figure 7 – This plot shows the variation of the mean of the range components with source orientation. The range components ($\Delta x$, $\Delta y$ and $\Delta z$) are shown on the abscissa. The ordinate shows the difference from the mean of the two perpendicular directions.

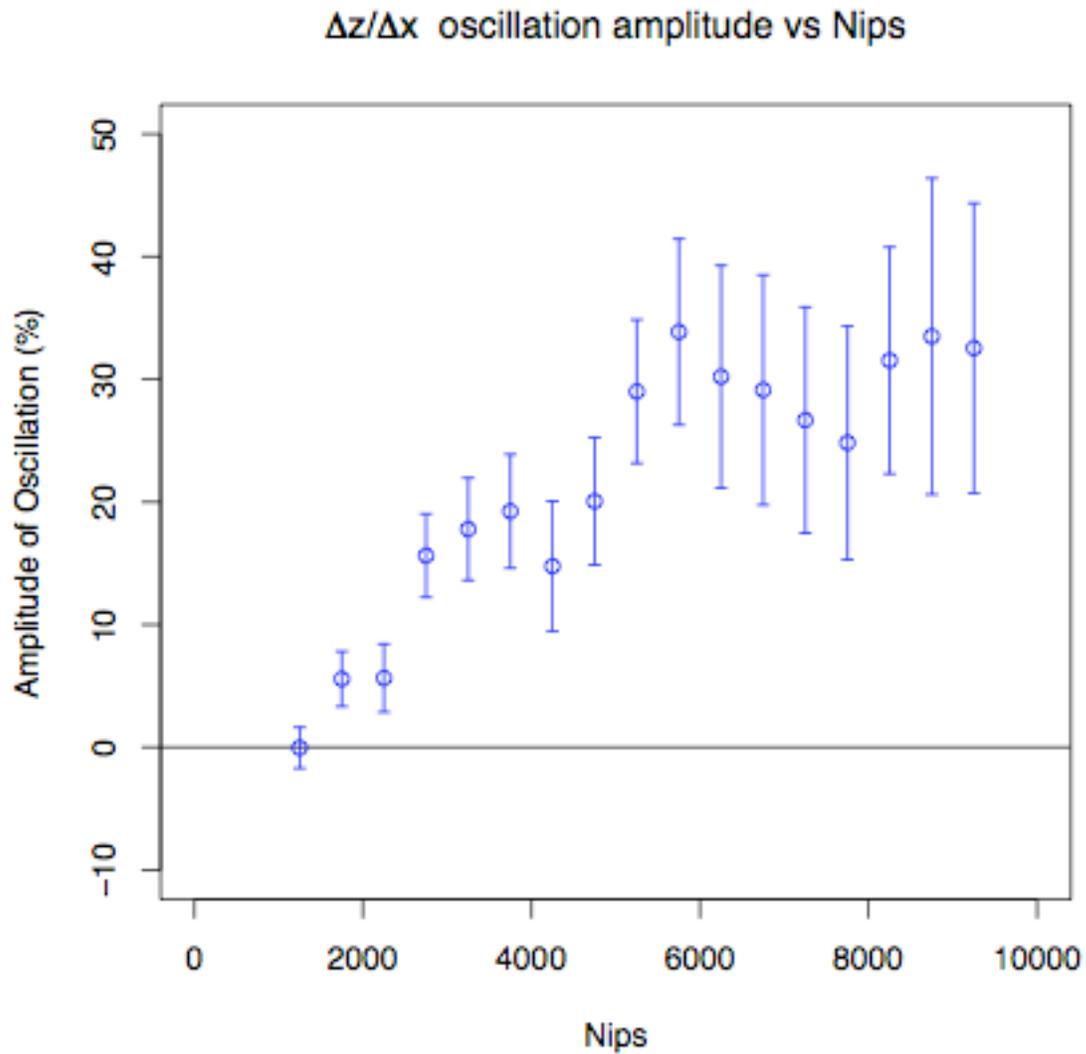

Figure 8 – This figure shows the amplitude of the *Δz/Δx* oscillation, expressed as a percentage of the average value, as a function of *NIPs* for 500 *NIPs* bins. The data show a strong correlation with *Nips*, or recoil energy, and also suggest that the effect disappears below the threshold of the detector ~1000 *NIPs*.